\newcommand{\beq}{\begin{equation}}
\newcommand{\eeq}{\end{equation}}

\newcommand{\lqc}{\Lambda_{QCD}}

\documentclass{ws-procs9x6}
\begin{document}

\title{Lower-dimension vacuum defects in lattice Yang-Mills theory}

\author{V.~I. ZAKHAROV}

\address{Max-Planck Institut f\"ur Physik \\
Werner-Heisenberg Institut \\ 
F\"ohringer Ring 6, 80805, Munich\\ 
E-mail: xxz@mppmu.mpg.de}


\maketitle

\abstracts{We overview lattice data on $d=1,2,3$ 
vacuum defects in four-dimensional gluodynamics.
In all the cases defects have total volume which scales in physical units
(with zero fractal dimension). In case of $d=1,2$ the defects are 
distinguished by ultraviolet divergent non-Abelian action as well.
This sensitivity to the ultraviolet scale allows to derive 
from the continuum theory strong
constraints on the properties of the defects. 
The constraints turn to be satisfied by the lattice data.
In the $SU(2)$ case we introduce 
a classification scheme of the defects which 
allows to (at least) visualize  the defect properties in a simple and unified
way. Not-yet-checked relation of the defects to the spontaneous
chiral symmetry breaking is suggested by the scheme.}

\section{Introduction}

Somehow, it went largely unnoticed that non-perturbative QCD has 
been changing fast.
The change is mostly due to results of lattice simulations
which ask sometimes for novel
continuum-theory models, see, in particular, \cite{simonov}.
Moreover, in many cases the lattice results 
are formulated in specific language and do not allow
for any immediate interpretation in terms of the continuum theory.
In this way, 
there arises a mismatch between richness of the lattice data
and scarceness of their interpretation  in the continuum theory.

An important example of this type are models of confinement.
Indeed, instantons still dominate thinking on non-perturbative
physics on the continuum-theory side. On the other hand, 
it is known from the lattice
measurements that the instantons do not confine \cite{confinement}. 
Moreover, the vacuum
fluctuations which are responsible for the confinement have been also
identified and turn to be monopoles and 
P-vortices, for review see, e.g.,\cite{review,greensite}.

There is no understanding whatsoever of these confining 
fluctuations in terms of the continuum theory.
Moreover, if now someone decides to go into  interpretation 
of the lattice data there is no regular way
to approach the problem. The point is that the 
monopoles and vortices are
defined on the lattice rather algorithmically, 
than directly in terms of
the gluonic fields. 
The central step is the
use of projections which replace, say, 
the original 
non-Abelian fields by the closest Abelian-field configurations.
The projection is a highly nonlocal procedure defined only
on the lattice
and blocks out any direct interpretation of the data.

To circumvent this difficulty, we attempt to summarize here the lattice
data on the confining vacuum fluctuations entirely in 
terms of continuum theory.
Hopefully, this could facilitate appreciation of the results 
of the lattice simulations. In particular, we emphasize that the confining
fluctuations appear to be vacuum defects of dimension lower than $d=4$.

What challenges the continuum theory in the most direct way  is a 
relatively recent discovery that the monopoles, see \cite{anatomy}
and references therein,  and vortices, see \cite{kovalenko}, are 
associated with ultraviolet divergent 
non-Abelian action. Since gluodynamics is well understood
at short distances, this newly discovered sensitivity of the
vacuum defects to the ultraviolet scale makes them subject 
to strong constraints
 from the continuum theory \cite{vz6}.
   
The presentation  is as follows. In Sect. 2 we summarize lattice data on
the confining fluctuations. In Sect. 3 
constraints from the continuum theory are 
outlined. In Sect. 4 possible relation to dual formulations of the Yang-Mills
theories is discussed.

\section{Lattice phenomenology}

\subsection{Total volume}

Imagine that indeed there exist low-dimension structures in the vacuum
state of gluodynamics.  Which $SU(2)$ invariants could be
 associated with such defects?
First of all, we could  expect that the total volume of the
corresponding defects scales in physical units.  What this means,
is easier to explain on particular  examples.

{\it $d=0$ defects}. In this  case, we are discussing density of points  in the
$d=4$ space. And the expectation for the total number
of the point-like defects would be
\beq\label{one}
N_{tot}~=~c_0\Lambda_{QCD}^4\cdot V_4~,
\end{equation}
where  $V_4$ is the  volume  of the lattice and $\Lambda_{QCD}$
can be understood either as a position of the pole of the (perturbative)
running coupling or, say,  as $\Lambda_{QCD}~=~\sqrt{\sigma_{SU(2)}}$
where  $\sigma_{SU(2)}$ is the string tension. Appearance of $\Lambda_{QCD}$
in (\ref{one}) would signal 
relevance  of the fluctuations to the confinement.

Equation (\ref{one}) could be readily understood if we were discussing
number of instantons.  However, it might worth mentioning from the very
beginning that the instantons actually do not belong to
the sequence of the vacuum fluctuations which we are going to consider.

{\it $d=1$ defects}. The $d=1$ defects are
lines. For the total length one can expect
\beq\label{two}
L_{tot}~=~c_1\Lambda_{QCD}^3\cdot V_4~~.
\end{equation}
Such defects can be identified with
the {\bf percolating monopoles}
(for latest data see
\cite{muller,boyko}). Percolation \footnote{For  theoretical background see, 
e.g., \cite{grimmelt}.}
means that 
there exists a large cluster of monopole trajectories
stretching itself through the whole volume of the lattice.
In the limit of infinite volume the percolating cluster also 
becomes infinite.
Note also that the monopole trajectories are closed by  definition, as a 
reflection of the monopole
charge conservation.

It is worth emphasizing that the scaling law
(\ref{two})  is  highly nontrivial
and, from the point of  view of the
lattice measurements, represents a spectacular phenomenon. 
 Indeed, on the lattice one changes arbitrarily the lattice spacing, 
$a$ while the corresponding bare coupling, $ g(a)$ is changed logarithmically,
according to the renormgroup. The scaling law (\ref{two})
implies that the probability 
$\theta(a)$ for a given link (actually, on the dual lattice)
to belong to the percolating cluster is changing as a power of $a$:
\beq\label{theta}
\theta (a)~\sim~ (\Lambda_{QCD}\cdot a)^3~~.
\end{equation}
Thus, in this case powers, and not logs, of the ultraviolet cut off
are observed.
In other words, there is no perturbative background
to  the defects which we are discussing and we are addressing
directly non-perturbative physics. 

{\it $d=2$ defects}. The defects are now two-dimensional surfaces and for
the total area the scaling law would read:
\beq\label{three}
  A_{tot}~=~c_2\Lambda_{QCD}^2\cdot~V_4~~.
\end{equation}
Such defects can be identified with {\bf percolating P-vortices}
which are known to satisfy (\ref{three}) \cite{greensite},
for the latest data see \cite{kovalenko}.  As in all other cases
discussed here the evidence is pure numerical, though. Because of the
space considerations we do not discuss here error bars,
concentrating only on the general picture. Details can be found
in the original papers.

{\it $d=3$ defects}. For a percolating three-dimensional volume we could expect:
\beq\label{four}
V_3~=~c_3\Lambda_{QCD}\cdot~V_4~~.
\end{equation}
First indications on existence of such defects were obtained recently \cite{volume}.

\subsection{Non-Abelian action associated with the defects}

Defects can be distinguished by their non-Abelian action as well. 
Since we have not specified yet the defects
dynamically, at first sight, we cannot
say anything on their action. Surprisingly enough, there exist educated guesses
concerning the non-Abelian action of  the defects
based on their dimension alone.  

{\it d=1 defects}. This case is singled out by 
the consideration that trajectories correspond to particles.
Particles, on the other hand, belong to field theory and we may hope to get insight 
into the properties of the trajectories from field theory. And, indeed, the action
\beq\label{classical}
S~=~M\cdot L~~,
\end{equation}
where $L$ is  the  length of the trajectory and $M$ is a mass  parameter
coincides with the classical action for a  free particle of mass $M$.
One may hope, therefore, that by evaluating 
 propagation of a particle as a  path integral with the action 
(\ref{classical}) 
one reconstructs the quantum propagator of a free particle.
And, indeed,
 this theoretical construction works. Moreover, it
is well known as the so called polymer approach to 
field theory, see, in particular,
\cite{sym}. Note that the use of the Euclidean (rather than Minkowskian) space
is  actually crucial to evaluate the corresponding path integral.
Also, one needs to introduce lattice to formulate the theory. 

Although the use of the action (\ref{classical}) does allow to recover the
free field  propagator, the {\bf propagating mass} turns to be not the
same $M$ but is  equal to
\beq\label{propagating}
m^2_{prop}~=~{(const)\over a}\big(M(a)~-~{\ln 7\over a}\big)~~,
\end{equation}
where the  constants $const, \ln 7$ are of pure geometrical origin and depend on
the lattice used. In particular, $\ln 7$ corresponds to  
the hypercubic lattice.

Note that in Eq (\ref{propagating}) we reserved now
for dependence of the mass parameter $M(a)$
on the lattice spacing $a$. Indeed, tuning of $M(a)$ to $\ln 7/a$ is needed
to get an observable mass (\ref{propagating})
independent on the
lattice spacing $a$.

Thus, our prediction for the action associated with $d=1$ defects (which are nothing else but the
monopole trajectories) is that the action is close to
\beq\label{ln7}
S_{mon}~\approx~{\ln 7\over a}\cdot L~~.
\end{equation}
Indeed, in this way we can explain that the length of the trajectories does not
depend on $a$, see Eq (\ref{two}).

Prediction (\ref{ln7})  does agree with the results of direct measurements of 
the non-Abelian action of
the monopoles \cite{anatomy}. Let us emphasize that the prediction (\ref{ln7})
does not use anything specific for monopoles and is rooted in the standard field theory.
Indeed, the polymer approach to field theory is no better no worse than other approaches.

{\it $d=2$ defects.} There exists simple theoretical argumentation in favor of an
ultraviolet  divergent action of the two-dimensional defects (or vortices) as well.
Consider the so called gluon condensate
\beq\label{condensate}
< ~(G^a_{\mu\nu})^2~>~\approx~{N_c^2-1\over a^4}\big(1~+ O(\alpha_s)\big)~~,
\end{equation}
where $G^a_{\mu\nu}$ is the non-Abelian field strength tensor and $a$ is the color
index. Note that the condensate (\ref{condensate}) on the lattice is in fact nothing else but
the average plaquette action \cite{gc}.

The vacuum expectation
value (\ref{condensate}) 
diverges as the fourth power of the ultraviolet cut off. This divergence is in 
one-to-one correspondence with the divergence in the density 
of the vacuum energy,
well known in the continuum theory. If one neglects interaction of the gluons,
the gluon condensate (\ref{condensate}) reduces to a sum over energies of 
the zero-point fluctuations.
That is why the r.h.s. of (\ref{condensate}) 
in the zero approximation is proportional to the number of degrees
freedom, that is to the number of gluons. 
Accounting for the gluon interaction brings in perturbative corrections.
For details  and most advanced calculations see
the last paper in Ref. \cite{gc}.

Note now that the monopole trajectories with the properties (\ref{two}) and (\ref{ln7})
give the following contribution to the the gluon condensate:
\beq\label{powerthree}
\langle ~(G^a_{\mu\nu})^2~\rangle _{mon}~\approx~{N_c^2-1\over a^4}(const)(\Lambda_{QCD}
\cdot a)^3~~.
\end{equation}
In other words, monopoles  correspond to a power like correction to the perturbative value
of the gluon condensate.

Now, the central point is that there exists rather well developed theory of the power
corrections, for review see
\cite{beneke}. The leading power-like correction is expected to be associated with
the so called ultraviolet renormalon. The corresponding contribution is of order
\beq\label{uv}
\langle ~(G^a_{\mu\nu})^2~\rangle _{uv-ren}~\approx~{N_c^2-1\over a^4}(const)(\Lambda_{QCD}
\cdot a)^2~~.
\end{equation}
Note that the power of $(\Lambda_{QCD}
\cdot a)$ here is different from what the monopoles  give, see Eq (\ref{powerthree}).

From this point of view it would  be unnatural  if the monopoles exhausted
the power corrections. Moreover, if the $d=2$ defects with the total area
satisfying (\ref{three}) have the action
\beq\label{areauv}
S_{vort}~=~(const){A_{vort}\over a^2}~~,
\end{equation}
then their contribution would fit the ultraviolet renormalon (\ref{uv}).
In this way one could have predicted (\ref{areauv}). The data
on the non-Abelian action associated with the P-vortices
\cite{kovalenko} do confirm the estimate (\ref{areauv}):
\beq\label{areauv1}
S_{vort}~\approx~0.54{A_{vort}\over a^2}~~,
\end{equation}
where $A_{vort}$  is the total area of the vortices.

{\it $d=3$ defects}. Proceeding to the three dimensional defects we could predict that the story
does  {\bf not} repeat itself and there is no ultraviolet action associated with the
$V_3$. Indeed, if the action would be of order $V_3/a^3$, the corresponding power
correction to the gluon condensate would exceed the ultraviolet renormalon, see (\ref{uv}),
by $(\Lambda_{QCD}\cdot a)^{-1}$ and contradict the theory.

The data \cite{volume} indeed do not indicate any excess of  the action
associated with the $V_3$. One can claim this to be a success of the theory.
It is also true, however, that the three dimensional defects so far are lacking identity
in terms of gauge invariant characteristics. 

\subsection{Entropy associated with the defects}

The ultraviolet divergences in the action of the monopoles,
see (\ref{ln7}), and of vortices, see (\ref{areauv}), suggest, at
first sight,  that these fluctuations are
not physical and exist only on scale of the lattice spacing $a$.
Indeed, the probability, say, to observe a monopole trajectory of length $L$ is
suppressed by the action as
\beq\label{entropy}
W(L)~\approx~\exp(-S_{mon})\cdot~(Entropy)~\sim~\exp(~-\ln 7 L/a)\cdot (Entropy)~~,
\end{equation} 
and in the continuum limit of $a\to 0$ the suppression due to the action is infinitely strong.
Since, on the other hand, the observed length of the monopole trajectories does not tend to zero
in the limit $a\to 0$ the suppression by action is to be canceled by the same strong
{\it enhancement} due to the entropy. Let us discuss this issue in more detail in cases
$d=1,2$.

{\it d=1 defects}. In this case, the entropy factor is  the number $N(L)$ of different
trajectories  of same length $L$. It is  quite obvious that for $L$ fixed and
$a\to 0$, $N(L)$ grows exponentially with $L$ \cite{polyakov2}. 
Indeed, trajectory on the lattice is
a sequence of steps of the length $a$. The number of steps is $L/a$ and at each step one 
can arbitrarily choose direction. The number of directions is determined by
the geometry of the lattice.
This is the origin of the  factor $\ln 7/a$ in Eq (\ref{propagating})
valid for the hypercubic lattice.

Therefore, the observation (\ref{ln7}) is nothing else but the statement
that in case  of the  monopole trajectories  the entropy is fine tuned to the action.
Moreover, since  there is actually no free parameter in the theory (the 
QCD coupling
is running and cannot be tuned) we are dealing rather with 
{\bf self tuning} of the monopole
trajectories. This should be a dynamical phenomenon.
 The author of the present review finds this observation of self-tuned objects on
the lattice absolutely
remarkable.

{\it d=2 effects}. The suppression due to the action is again there, see (\ref{areauv1}).
Moreover, from lattice measurements one can conclude that the action is not simply proportional
to the total area of the vortex but depends also on local geometrical properties of the
surface, for detail see \cite{syritsyn}.

As for the entropy, the lattice measurements indicate \cite{syritsyn} that
mutual orientation of adjacent plaquettes belonging to the
vortices is random. This observation, in turn, implies exponential enhancement of the
entropy of the surfaces.

Theoretically, one can determine (from indirect argument) entropy of P-vortices in case of
$Z_2$ gauge theory. In case of the $Z_2$ theory plaquettes take on values $\pm 1$ and 
the P-vortices by definition pierce the negative plaquettes.
Numerically,
\beq\label{entropyz2}
(Entropy)_{Z_2-vortices}~\approx~\exp(~0.88 A_{tot}/a^2)~~,
\end{equation}
where $A_{tot}$ is the total area of the negative plaquettes.

We see that the action (\ref{areauv1}) is {\it not} tuned to the entropy
of $Z_2$-vortices (\ref{entropyz2}). A straightforward interpretation of this observation is that
the entropy of the P-vortices in SU(2) gluodynamics is lower than that
in case of the $Z_2$ theory. As we will discuss later, the P-vortices  in the gluodynamics case
are populated by magnetic monopoles. The lower value of the entropy could be understood as evidence
that at short distances it is rather monopoles which are to be considered as fundamental
while the P-vortices `follow' the monopole trajectories. This issue deserves further detailed 
study and discussion.

\subsection{Alignment of geometry and non-Abelian fields}

 So far we considered geometry and non-Abelian fields associated with the 
defects separately. The total volume is a geometrical scalar and the total
action is a scalar constructed on the fields. Of course, 
observation of invariant characteristics which mix up 
geometry and fields would be even more  interesting.

{\it d=3 defects.} Let us start with $d=3$ and assume existence of three dimensional
volumes to be granted. The volume percolates on the scale $\Lambda_{QCD}^{-1}$.
However, if we tend $a\to 0$ then we can think in terms of large 
(in the lattice  units)
$d=3$ defects. What are possible $SU(2)$ invariants associated with these
volumes?

In $d=3$ the $SU(2)$ gluodynamics is described by three  vectors ${\bf H}^a$ (a=1,2,3)
where the `magnetic fields' ${\bf H}^a$ are vectors both in the coordinate and color spaces.
Generically, there are three independent vectors and the simplest invariant constructed on
these fields is their determinant. However, the absolute value of the determinant
varies from one point of the $d=3$ volume to another. The invariant which can be associated
with the whole volume is, obviously, the sign of the determinant 
\footnote{The classification scheme of the defects we are proposing below, to our knowledge,
has not been discussed in literature
(an earlier version can be found in  Ref. \cite{feodor}). However, at least partially, it is close to or
motivated by well known papers, see, in particular, \cite{thooft1}.
It is worth emphasizing that we are using the continuum-theory language,
assuming fields to be continuous functions of the coordinates.  Since on the lattice
the measurements are performed on the scale of the lattice spacing
$a$, the lattice fields fluctuate wildly on the same scale. Thus, the underlying
assumption is that the continuum-theory language is still valid on average,
so that the perturbative fluctuations do not interfere with topology.}
 :
\beq\label{I3}
I_3~=~sign \{\epsilon^{ikl}\epsilon_{abc}H^a_iH^b_kH^c_l\}~~.
\end{equation}
Unfortunately, there are no lattice data which could confirm or reject 
this prediction.

{\it $d=2$ defects}. The invariant (\ref{I3}) is uniquely defined as far as 
all three magnetic fields are indeed independent.
If there are only two independent vectors the determinant has zero of first order
and geometrically we have then a closed $d=2$ surface, as a boundary of the $d=3$ defects.
It seems natural to speculate that these boundaries are our $d=2$ defects, or vortices.

Thus, we come to the prediction that 
the percolating $d=2$ surfaces and non-Abelian fields are 
aligned with each other. In other words, the non-Abelian fields associated
with the vortices and resulting in  the excess of the action (\ref{areauv})
are predicted to spread over the surface while the perpendicular component
is to vanish. This prediction works \cite{kovalenko} {\it perfectly well}.

{\it $d=1$ defects}.
The next step is an reiteration
of the previous one. Namely, there could be  
zeros of the second order of the determinant (\ref{I3}) 
and geometrically 
zeros of the second order are  closed
trajectories.  Moreover, if  these closed lines are monopole 
trajectories (which are indeed closed by definition), 
the non-Abelian field of the monopoles is to be aligned with
their trajectories. This expectation {\it agrees } with
the existing data \cite{giedt} as well.

It is amusing that the monopoles are expected to be locally Abelian.
Indeed, there is only one independent color magnetic field associated with 
zeros of second order of the determinant (\ref{I3}).
Moreover, they are singular, see (\ref{ln7}).
Thus, we are coming to an after-the-fact justification of
the use of the Abelian projection to detect the monopoles.
On the other hand, their field is absolutely not spherically symmetrical
(which would be the case for the Dirac monopoles). And this spatial asymmetry
is manifested in the measurements \cite{giedt}.

\subsection{Spontaneous breaking of chiral symmetry}

Note that our classification scheme predicts that the $d=3$ defects are
characterized by invariants which distinguish between left- and right-
hand coordinates. Of course, on average the regions with determinants
(\ref{I3}) positive and negative occupy the same volume.
(Moreover, in the continuum limit the the
percolating $d=3$ volume, see Eq (\ref{four}),  
occupies a vanishing part of the whole  $d=4$ space).

As  is mentioned above there are no direct measurements
of the sign of the determinant (\ref{I3}) within the d=3 defects.
However, it is known that removal of the P-vortices
restores chiral symmetry \cite{chiral}. Also,
there exists  independent evidence  that some 3d volumes 
are related to the spontaneous symmetry breaking 
in case of $SU(3)$ gluodynamics \cite{horvath} 
\footnote{An obvious generalization of our invariant (\ref{I3})
to the SU(3) case would be $sign~\{f_{abc}\epsilon^{ikl}H^a_iH^b_kH^c_l\}$.}.
However, it seems too early to speculate that 
the two 3d volumes in question are actually the same.

\section{Theoretical constraints}

\subsection{Consistency with the asymptotic freedom}
 
It is worth emphasizing that there is no developed theory
of the defects considered. So far we summarized
lattice observations. Moreover,  we have been even avoiding discussion
how  the defects are defined and observed on the lattice,
for reviews see, e.g., \cite{review,greensite,vz5}. Let us only mention that
the guiding principle to define the defects was the search
for effective infrared degrees of freedom responsible
for the confinement.

However, what is most amusing from the theoretical point of view
is that the defects have highly non-trivial
properties in the ultraviolet. 
The ultraviolet divergence in the action,
see (\ref{ln7}), (\ref{areauv1}) are most remarkable.

Indeed, Yang-Mills theories are well understood at short distances.
The only divergence 
which is allowed on the fundamental level
is that one in the coupling $\alpha_s$.  Thus, one could argue
that all the ultraviolet divergences are calculable perturbatively
in asymptotically free theories. Moreover, this statement seems rather trivial.
What is actually  not so trivial is that on the lattice one can also consider
power-like divergences, see, e.g., (\ref{powerthree}), (\ref{uv}).
In the continuum theory, power-like ultraviolet divergences usually are used,
at best, for estimates. On the lattice, the ultraviolet cut off is introduced 
explicitly and one can treat
 power-like divergent observables in a fully quantitative way,
see, e.g., (\ref{condensate}).  This extends in fact the predictive power of the theory.

It is, therefore, no surprise that using the asymptotic freedom 
one can derive strong constraints
on the properties of the vortices \cite{vz6}. 

\subsection{Classical condensate $<\phi_{magn}>$}

Let us consider first
the vacuum condensate of the magnetically charged field $<0|\phi_{magn}|0>$.
Of course, in the YM theory there is no fundamental magnetically charged
field. However, the monopole trajectories are observed on
the lattice.  Using the polymer approach to field theory 
we can translate the lattice data on trajectories into a field-theoretic language,
see Ref \cite{maxim,vz6} and references therein.
The only assumption is that there exists an effective field theory for
the magnetically charge field.

In particular, the percolating cluster corresponds to the
classical vacuum expectation value $<\phi>$.
One can derive \cite{maxim}:
\beq\label{classical1}
<\phi_{magn} >^2~\approx~{a\over 8}\rho_{perc}~\approx~(const)\Lambda^2_{QCD}(a\cdot \Lambda_{QCD})~.
\end{equation}

\subsection{Vacuum expectation value $<|\phi_{magn}|^2>$}

One can also expect
that there exist quantum fluctuations. And indeed, apart from the percolating cluster,
there observed finite monopole clusters. 
Which are naturally identified with the quantum fluctuations.
A basic characteristic for these clusters
 is again their total length.
By definition:
\beq
L_{tot}~\equiv~L_{perc}~+~L_{fin}~\equiv~\rho_{perc}\cdot V_4~+~\rho_{fin}\cdot V_4~~,
\end{equation}
where $\rho_{perc}, \rho_{fin}$ are called the densities of the 
percolating and finite monopole clusters, respectively.

Using the polymer approach to the field theory one can express
the vacuum expectation value of the magnetically charged field
in terms of the  monopole trajectories \cite{maxim}:
\beq\label{phi2}
< ~|\phi|^2~>~=~{a\over 8}\big(\rho_{perc}~+~\rho_{fin}\big)~~.
\end{equation}
Instead of deriving this relation (which is also quite straightforward)
let us explain why (\ref{phi2}) is natural.
Concentrate on the quantum fluctuations, that is on $\rho_{fin}$.
Moreover, consider small clusters $L<<\Lambda_{QCD}^{-1}$.
Then there is no mass parameter at all and on pure dimensional
ground one would expect:
\beq\label{dimension}
\big(\rho_{fin}\big)_{dimension}~=~{const\over a^{d-1}}~=~{const\over a^3}
\end{equation}
where $d$ is the number of dimensions of space and we consider the $d=4$ case.
If this dimensional estimate held, then the vacuum expectation value would be
quadratically divergent in the ultraviolet:
\beq\label{quadratic}
<~|\phi|^2~>_{dimension}~\sim ~a^{-2}~~.
\end{equation}
And we rederive the standard quadratic divergence in the vacuum expectation
value  of a scalar field.  

Now, the central point is that although we call all these estimates
`dimensional' or `natural'  we are not allowed to have (\ref{quadratic}).
Indeed, Eq. (\ref{quadratic}) would hold for an {\it elementary} scalar field.
However, we are not allowed to have new elementary particles at short distances.
Because of the asymptotic freedom, there are only free gluons at short distances.
And what we are allowed to have for the vacuum expectation value in point?
Clearly:
\beq\label{allowed}
<~|\phi|^2~>_{allowed}~\sim ~\Lambda_{QCD}^2~~.
\end{equation}
In terms of the monopole trajectories (which are our observables) Eq (\ref{allowed})
reduces to:
\beq\label{allowed1}
\rho_{fin}~\sim ~{\Lambda^2_{QCD}\over a}~~.
\end{equation}
It is most remarkable that the data \cite{muller,boyko} do comply with
(\ref{allowed1})!

\subsection{Branes}

It is of course very gratifying that the data comply with the constraint
(\ref{allowed}). On the other hand, the reader may feel that our
summary of the phenomenology looks 
self contradictory. Indeed, first we observed that the non-Abelian monopole
action corresponds to a point-like particle, see  
(\ref{ln7}). But then  we said that there should be no
new particles, and the data agree with that constraint.

Still, there is no contradiction between these observations. Rather, 
taken together they amount to observation of a new object, which can be
called branes. Indeed, cancellation of $\ln 7/a$ in the equation
for the mass is needed to balance the entropy at very short distances
of order $a$. Eq (\ref{allowed}), on the other hand, is a global constraint.
The geometrical meaning of this constraint is actually transparent. 
Namely, it means that on large scale the monopoles live not on the whole
$d=4$ space but on its $d=2$ subspace. 

Within the classification scheme of the defects which we discussed above,
this association of the monopoles with surfaces is automatic.
Algorithmically, however, the trajectories and surfaces are defined
independently. And the fact that the monopole trajectories do belong to surfaces 
is highly non-trivial from the observational point of view.

Thus, what is observed on the lattice are $d=2$ surfaces populated with ``particles''
(better to say with the tachyonic mode of the monopole field).
When we call this objects branes we emphasize that the affinity
of the monopoles to the surfaces remains  true even at the scale $a$.
Traditional discussions of the P-vortices and monopoles emphasize,
on the other hand, physics in infrared, 
and one talks about `thick vortices', see, 
e.g. \cite{greensite}.  

\subsection{Implications for models}

{\it Abelian Higgs model.} The lattice data on monopoles are usually 
interpreted in terms of
an effective Abelian Higgs model, see in particular
\cite{suzuki} and references therein. Our Eq (\ref{classical1})
implies, however,
\beq
<\phi_{magn}>~\sim~(a\cdot \Lambda_{QCD})^{1/2}\Lambda_{QCD}~
\end{equation}
It is most remarkable that the classical condensate vanishes in
the continuum limit $a\to 0$. 
Nevertheless, the heavy quark potential 
at large distances generated by the monopoles remains the same
since it is determined entirely by $\rho_{perc}$ 
which scales in the physical units!

To establish a relation to the standard fit to the Abelian Higgs model
one should use, most probably, matching of the two approaches at some
$a$ \footnote{The author would like to acknowledge discussion of this
point with M.N. Chernodub}.

{\it Gauge invariant condensate of dimension two.}
Eqs (\ref{allowed}), (\ref{allowed1}) provide us with a value for a gauge 
invariant
condensate of dimension two:
\beq\label{twonumber}
<~|\phi_{magn}|^2~>~\approx~a\cdot\rho_{fin}~\sim~\Lambda_{QCD}^2~.
\end{equation}
In terms of the fundamental variables, condensate of dimension two was
introduced \cite{two} as the minimum value along the gauge orbit
of the gauge potential squared
$(A^a_{\mu})^2_{min}$  . While the $<(A_{\mu}^a)^2>_{min}$
is contaminated with  perturbative divergences, the condensate (\ref{twonumber})
provides us, on the phenomenological level, directly with a non-perturbative
condensate of dimension two.
Existence of a gauge invariant condensate of dimension two is crucial
for models of hadrons, see, in particular, \cite{niemi}.
On theoretical side, the nonperturbative part of the condensate of dimension two
in the Hamiltonian picture
is related to the Gribov horizon \cite{vanbaal}.

{\it Percolation.}
The fact that the monopole action is ultraviolet divergent, see (\ref{ln7}),
allows to consider them as point-like and apply percolation theory. 
In particular, one derives in this way the spectrum of finite
monopole clusters as function of their length $L_{fin}$:
\beq
N(L_{fin})~\sim {1\over L_{fin}^3} ~~,
\end{equation}
which agrees with the data, for details see \cite{maxim}.

Moreover, it is a common feature of percolating systems that
near the phase transition, in the supercritical phase, the density of
the percolating cluster is small. Then the observation (\ref{theta})
can be interpreted as indication that in the limit $a\to 0$ we hit
the point of second-order phase transition. 

Note also that second order phase transition is associated usually with
a massless excitation. At first sight, there is no massless excitation
in our case, however. The resolution of the paradox is that percolation
(or randomization) happens now on the ultraviolet scale. Respectively,
all the masses are to be measured in the lattice units, $1/a$. In this sense
a glueball mass,
\beq
m^2_{glueball}~\sim~{1\over a^2}\cdot (a\cdot\Lambda_{QCD})^2~~,
\end{equation}
in the limit $a\to 0$ corresponds to a massless excitation. 
Moreover, one expects that the glueball mass controls the scale of fluctuations of
the monopole clusters. But these expectations have not been checked yet
(see, however, \cite{ishiguro}).

{\it Singular fields.}  Although we are repeatedly emphasizing 
that the branes are associated
with singular fields it should also be mentioned that the 
non-Abelian fields considered are 
significantly smaller than the corresponding projected fields.
In particular, the Dirac monopole would have a larger action:
\beq   
S_{Dir}~\sim~{1\over g^2}\cdot {L\over a}~\gg~\ln 7{L\over a}
\end{equation}
where $g^2$ is the gauge coupling and we consider the limit $g^2\to 0$.
Already at presently available $g^2$ the action calculated in terms of
the projected Abelian fields (corresponding to the Dirac monopole)
is a few times larger than the actual non-Abelian action which determines
the dynamics and which we discussed so far. The same is true for the vortices.

This distinction  -- in terms of the action -- 
between the Dirac and lattice monopoles
is very important from theoretical point of view. 
The observed monopoles are associated
with singular non-Abelian fields but these fields 
are no more singular than ordinary
zero-point fluctuations, or perturbative fields. 
The Dirac monopoles, on the other hand,
in the limit of $g^2\to 0$ would be more singular than the perturbative fields.
According to the standard ideas of the lattice theories, such fields could actually
be removed without affecting the basic 
physical content of the theory. The lattice monopoles and vortices, on the other
hand, cannot be removed without affecting perturbative fluctuations as well. In the
next two sections we will consider this issue in more detail.

 \section{Towards duality}

It is a well known  that topological excitations of a `direct' formulation
of a theory may become fundamental variables of a dual formulation
of the same theory. Examples can be found, e.g., in the review \cite{savit}.
Little, if anything, known theoretically on the dual formulation of
the Yang-Mills theories without supersymmetry.
Nevertheless, generically one might think in terms of branes \cite{maldacena}.
In case of supersymmetric extensions of YM theories the branes are
classical solutions. One could speculate that the branes discussed in this review
are `quantum branes'. 
Of course, it remains a pure speculation until
something definite could be said on the properties of the quantum branes on the
theoretical side. It is amusing, however, that there is a sign of duality
between the branes discussed in the preceding section and high orders of perturbation
theory \cite{vz8}.  

\section{Long perturbative series}

\subsection{Expectations}

Let us start with reminding the reader some basic facts about 
perturbative expansions, for detailed reviews see, e.g., \cite{beneke}.
A generic perturbative expansion for a matrix element of a local operator
 looks as:
\beq \label{pert}
\langle~ O~\rangle~= ~(parton~model)\cdot
\big(1~+~\sum_{n=1}^{\infty}a_n\alpha_s^n~\big)~~,
\end{equation}
where we normalized
the anomalous dimension of the operator $O$ to zero
and $\alpha_s$ is small,
$\alpha_s\ll 1$.

In fact, expansions (\ref{pert}) are only formal
since the coefficients $a_n$ grow factorially at large $n$:
\beq\label{growth}
|a_n|~\sim~c_i^n\cdot n!~~,
\end{equation}
where $c_i$ are constants. Moreover, there are a few sources
of the growth (\ref{growth}) and, respectively, $c_i$
can take on various values.
The factorial growth of $a_n$ implies that the expansion (\ref{pert})
is asymptotic at best. Which means, in turn, 
that (\ref{pert}) cannot approximate
a physical quantity to accuracy better than
\beq\label{uncert}
\Delta~\sim~\exp\big(-1/c_i\alpha_s\big)~\sim~
\Big({\lqc^2\cdot a^2}\Big)^{b_0/c_i}~~,
\end{equation}
where $b_0$ is the first coefficient in the $\beta$-function.
To compensate for these intrinsic uncertainties one modifies
the original expansion (\ref{pert}) by adding the corresponding
power corrections
with unknown coefficients.

In case of the gluon condensate the theoretical expectations
can be summarized as:
\beq\label{expectations}
\langle 0|~{-\beta(\alpha_s)\over\alpha_s b_0}
\big(G_{\mu\nu}^a\big)^2|0\rangle~
\approx~\alpha_s{(N_c^2-1)\over a^4}\big(1+
\sum_{n=1}^{\sim N_{ir}}a_n\alpha_s^n~+~(const)a^4\cdot\Lambda_{QCD}^4\big)~~,
\end{equation}
where
$$N_{ir}~\approx~{2\over b_0\alpha_s}~~$$ 
and terms proportional to $\Lambda_{QCD}^4$ correspond to  
$<0|(G_{\mu\nu}^a)^2|0>_{soft}$ which enters the QCD sum rules, 
for review see \cite{sumrules}.
 
A conspicuous feature of the prediction (\ref{expectations})
is the absence of a quadratic correction,
compare (\ref{uv}).
Thus, we are seemingly coming 
to a contradiction between the lattice branes and
continuum-theory perturbation theory. 
Let us, however, have a closer look at the problem.

\subsection{Numerical results}

Numerically, this perturbative expansion for the gluon condensate 
was studied in greatest detail  in
the papers in Ref \cite{gc} (especially in the latest one). 
The results can be summarized in the following way.
On the lattice, the gluon condensate is nothing else but
the average plaquette action.
Represent the plaquette action $\langle P\rangle $ as:
\beq\label{plaquette}
a^4\langle P\rangle ~\approx~P_{pert}^N~+~b_Na^2\lqc^2~+c_Na^4\lqc^4~~,
\end{equation}
where the average plaquette action $\langle P\rangle $
is measurable directly on the lattice and is known to high accuracy,
$P_{pert}^N$ is the perturbative contribution calculated 
up to order N:
\beq\label{pn}
P_{pert}^N~\equiv~1~-~\sum_{n=1}^{n=N}p_ng^{2n}~~,\end{equation}
and, finally coefficients $b_N,c_N$ are fitting parameters
whose value depends on the number of loops $N$. Moreover, the form of
the fitting function (\ref{plaquette})
is rather suggested by the data 
than imposed because of theoretical considerations.

The conclusion is that up to ten loops, $N=10$ it is the quadratic correction
which is seen on the plots while $c_N$ are consistent with zero.
However, the value of $b_N$ decreases monotonically with growing $N$ 
\cite{gc}. The factorial divergence (\ref{growth})
is not seen yet and perturbative series reproduces
the measured plaquette action at the level of $10^{-3}$. 
Finally, at the level $10^{-4}$  the $\lqc^4$ term seems to 
emerge \cite{gc}.
 
\subsection{Implications}

Thus, there is a fundamental difference between the instantons and branes.
The instantons correspond to the `soft' gluon condensate and are hidden
in the $(\Lambda_{QCD}\cdot a)^4$ corrections which are not calculable
perturbatively. In short, instantons are added to perturbation theory.

The branes, on the other hand, appear to be dual to long perturbative series.
If one is able to calculate many orders of perturbation theory, there
is no need to account for the branes as far as {\it local} quantities are
concerned.

This might first sound disappointing for those who is beginning to believe
in the important role of the lattice branes. In fact 
it is not disappointing at all.
To the contrary, we have first firm piece of evidence that the branes belong to a 
dual world. To reiterate: the instantons belong to the `direct' formulation and they are added
to the perturbation theory of the direct formulation. Branes belong to the dual formulation.
Adding them to the perturbation theory of the direct formulation would be mixing up 
two different (dual to each other) representations of the same theory.

Actually, the very existence of  fine tuned branes could not be understood
within the direct formulation of the YM theory. However, if there were an
existence theorem for a dual formulation, the fine (or self-) tuning was
implied by the theorem. Reversing the argument, we can say that observation of the
fine tuning might indicate existence of a dual formulation.
 
\section*{Acknowledgments}

This mini review is written on the occasion of the seventieth birthday of
Yuri Antonovich Simonov. I am thankful to him for
many years of sharing his ideas on QCD, discussions and friendship.

This review is devoted to interpretation of the lattice
data obtained mostly by colleagues from ITEP:
P.Yu.  Boyko,
M.N. Chernodub, F.V. Gubarev, A.N. Kovalenko, M.I. Polikarpov, S.N. Syritsyn.
I am thankful to all of them for common work, used in these notes,
and numerous discussions. 
I am thankful to R. Alkofer, A. DiGiacomo, I. Horvath,
Ch. Gattringer, J. Greensite, L. Stodolsky, T. Suzuki
for enlightening  discussions.
  
The notes above were presented as a talk at the Workshop
on quantum field theory at Ringberg in February 2004.
The author is thankful to the organizers, W. Zimmermann,
D. Maison, E. Seiler for the invitation and hospitality,
 
The text of the review was finalized while 
the author was visiting Pisa University.
I am thankful to A. DiGiacomo for the hospitality. 
The work was partially supported by the grant INTAS-00-00111 and
and by DFG program ``From lattices to phenomenology of hadrons''.

\end{document}